\documentclass[a4paper,UKenglish]{lipics}

 \usepackage{microtype}%if unwanted, comment out or use option "draft"

\newcommand{\J}{\mathcal{J}}
\newcommand{\G}{\Gamma}
\newcommand{\Ge}{\mathcal{G}}
\newcommand{\Sc}{\mathcal{S}}
\newcommand{\fa}{\forall}
\newcommand{\ex}{\exists}
\newcommand{\eqdef}{=_{def}}

\title{How to prove it in Natural Deduction: A Tactical Approach
\footnote{This research has been supported by project UNAM-PAPIIT (IN400514-3).}
}
\author[1]{Favio E. Miranda-Perea}
\author[1]{P. Selene Linares-Arévalo}
\author[2]{Atocha Aliseda}
\affil[1]{ Departamento de Matemáticas, Facultad de Ciencias, Universidad Nacional Autónoma de México. 
Circuito Exterior s/n, Ciudad Universitaria C.P 04510, México D.F. \\
  \texttt{favio@ciencias.unam.mx  \quad   selene\_linares@ciencias.unam.mx}}
\affil[2]{Instituto de investigaciones Filosóficas. Universidad Nacional Autónoma de México. 
  Circuito Mario de la Cueva s/n, Ciudad Universitaria, C.P 04510, México D.F. \\
  \texttt{atocha@filosoficas.unam.mx}}

\authorrunning{F.\,E. Miranda-Perea , P. Selene Linares-Arévalo and A.\, Aliseda} 

\Copyright{Favio E. Miranda-Perea , P. Selene Linares-Ar\'evalo \\ \phantom{XXXXXXX} 
and Atocha Aliseda}

\serieslogo{logo_ttl}%please provide filename (without suffix)
\volumeinfo%(easychair interface)
  {M. Antonia {Huertas}, Jo\~ao {Marcos}, Mar\'ia {Manzano}, Sophie {Pinchinat}, \\
  Fran\c{c}ois {Schwarzentruber}}% editors
  {5}% number of editors: 1, 2, ....
  {4th International Conference on Tools for Teaching Logic}% event
  {1}% volume
  {1}% issue
  {157}% starting page number
\EventShortName{TTL2015}
\begin{document}

\maketitle

\begin{abstract}
The motivation for this paper comes out of our experience with teaching natural deduction (ND) and with the way this formal system is implemented by the {\sc Coq} proof assistant, namely by means of so-called tactics, which are heuristics that transform a goal formula into a sequence of subgoals whose provability implies that of the original formula. We aim at capturing some of these tactics into a system of ND for minimal logic. Our goal is twofold: formal and didactic. The former delivers a formal system with its underlying heuristics to build proofs, which in turn serves our latter purpose, that of making an ideal system for the teaching of ND at an undergraduate level in a computer science program.
 \end{abstract}

\section{Introduction}
The importance of logic in mathematics and computer science is unquestionable. The use of proof-assistants, whose kernel are implemented logics, in the verification and certification of software and mathematical proofs is coming of age. A proof-assistant is a computer system that consists of a domain-specific language representing logical objects, as well as definitions and theorems about these objects, together with a mechanism that allows for the interactive construction and validation of proofs. Though proof assistants have been used to teach logic (e.g. \cite{pw}), we think there is still a gap between the traditional way of teaching deductive systems, in particular natural deduction (ND from now onwards), using \cite{hr} for example, and the use of proof-assistants to solve 
tasks of software construction and verification, like the ones tackled in a programming language foundations course 
(c.f. \cite{ppl}). This paper claims to be a contribution to filling this gap. 

Our main goal is to teach ND to undergraduate computer science students in the way this formal system is implemented by a proof-assistant, specifically {\sc Coq} (\url{http://coq.inria.fr}). This way, the migration from studying logic, to using it, for example in a programming languages foundations or a formal verification course, will be smooth, as the transition will be mainly syntactical. We pursuit this goal from the theoretical side, that is, we do not discuss here how to use {\sc Coq} to teach logic. To the best of our knowledge, our approach is novel, for we capture {\sc Coq}'s native proof search mechanism for first order logic by giving a formal definition of its procedural way of constructing proofs, which corresponds to what we may call ``usual mathematical reasoning''.

\section{Natural Deduction}
ND is, as the name says, a formalism that captures natural reasoning, as opposed to the formal reasoning given by Hilbert's axiomatic systems, recall for example the derivation of $A\to A$ in such a system.  The flavor of ND we choose is one with localized hypotheses, called S-systems in \cite{indr}, for this version is more akin to computer science applications. In these systems, inference rules are not applied over formulas, but over judgments (sequents), which are pairs of the form $\G\vdash A$ where $A$ is a formula and $\G$ is a set of formulas called the context.
The intuitive meaning of such a sequent is that formula $A$ is provable from the hypotheses in $\G$. This way, all current hypotheses, i.e. formulae in $\G$,  are available at every derivation step. Therefore there is no need for a global discharging mechanism\footnote{i.e. One that involves specific previous parts of the derivation.}, unlike Fitch-style ND as presented in \cite{hr,mh}. The specific inference rules are: 
\[
\mspace{275mu} \frac{}{\G,A\,\vdash A}\;(Hyp)
\]

\[
\mspace{120mu} \frac{\Gamma,A\vdash  B}{\Gamma\vdash A\rightarrow  B}\;(\rightarrow\! I) \qquad \qquad \qquad
\frac{\Gamma\vdash A\rightarrow  B\;\;\;\Gamma\vdash A}{\Gamma\vdash  B}\;(\rightarrow E)
\]

\[
\mspace{60mu} \frac{\G\vdash A\;\;\;\;\;\G\vdash B}{\G\vdash A\land
   B}\;(\land I) \qquad \quad
\frac{\G\vdash A\land B}{\G\vdash A}\;(\land E)\qquad \quad
\frac{\G\vdash A\land B}{\G\vdash B}\;(\land E)
\]

\[
\mspace{165mu} \frac{\G\vdash A}{\G\vdash A\lor B}\;(\lor I)\qquad \qquad \qquad
\frac{\G\vdash B}{\G\vdash A\lor B}\;(\lor I)
\]

\[ 
\mspace{190mu}\frac{\G\vdash A\lor B\;\;\;\;\;\G,A\vdash
  C\;\;\;\;\;\G, B\vdash C}
{\G\vdash C}\;(\lor E)\;\;
\]

\[
\mspace{150mu} \frac{\G\vdash A\;\; x\notin FV(\G)}{\G\vdash \fa x A}(\fa I) \qquad \qquad
\frac{\G\vdash \fa x A}{\G\vdash A[x:=t]}(\fa E)
\]

\[
\mspace{70mu}\frac{\G\vdash A[x:=t]}{\G\vdash \ex x A}(\ex I) \qquad \qquad
\frac{\G\vdash \ex x A\;\;\;\;\;\G,\, A\vdash C \;\;\;\;\;x\notin FV(\G,C)}
{\G\vdash C}(\ex E)
\]

As usual, we have a starting rule or axiom scheme (Hyp); introduction rules (I), useful to prove a formula according to its logical form (i.e., its main connective or quantifier) and elimination rules (E), useful to obtain information from a formula. Note that neither the false constant $\bot$
 nor the negation operator $\neg$ are present and therefore we are dealing
with minimal logic. The reason is that we are interested in logic from an
algorithmic point of view and classical logic departs from this
view. Our notion of proof is therefore linear as opposed to a tree form.
From a theoretical perspective, this choice is irrelevant, but the linear
form is in accord with the way proof-assistants operate.

\begin{definition}[]\label{def:deriv}
A proof or derivation of judgement $\mathcal{J}=_{def}\G\vdash A$ from the
set of judgements $\mathbb{H}$ is a finite sequence of
judgements $\Pi=\langle \J_1,\ldots,\J_k \rangle$ such that $\J_k=\J$ and for every $1\leq i\leq k$ one of the following conditions hold:
\begin{itemize}
\item $\J_i$ is an instance of the $(Hyp)$ rule
\item $\J_i\in\mathbb{H}$
\item For every $1\leq i \leq k,\;\J_i$ is the conclusion of an instance of some inference rule whose premises are $\J_{l_1},\ldots,\J_{l_n}$ with $l_1,\ldots,l_n< i$
\end{itemize}

We say that $\G\vdash A$ is provable, or simply that $\G\vdash A$ holds, if it is provable from the empty set of sequents $\mathbb{H}=\varnothing$.
\end{definition}

\begin{example}\label{ex:wex}
Let $\G=\{p \to q \lor r \; , \; q\to r \; , \; r\to s\}$,  we want to show that $\G\vdash p\to s$ holds. The following is a derivation of this 
sequent\footnote{\noindent The last column is not part of the formal derivation, but 
consists of justifications for every step, according to definition \ref{def:deriv}.}: 
\[
\begin{array}{rll}
1 \quad & \G,\;p  \; \vdash \; p & \qquad (Hyp) \\
2 \quad & \G,\;p  \; \vdash \; p\to q\lor r & \qquad (Hyp) \\
3 \quad & \G,\;p  \; \vdash \; q\lor r & \qquad  (\to E) \; 1,2 \\
4 \quad & \G,\;p,\;q \vdash \; q & \qquad (Hyp) \\
5 \quad & \G,\;p,\;q \vdash \; q\to r & \qquad (Hyp) \\
6 \quad & \G,\;p,\;q\vdash\; r & \qquad (\to E) \; 4,5 \\
7 \quad & \G,\;p,\;r\vdash\; r & \qquad (Hyp) \\
8 \quad & \G,\;p \vdash\; r & \qquad (\lor E) \; 3,6,7  \\
9 \quad & \G,\;p\; \vdash \; r\to s & \qquad (Hyp)  \\
10 \quad & \G,\;p\; \vdash \; s & \qquad (\to E) \; 8,9  \\
11 \quad & \G\;\vdash \; p\to s &\qquad (\to I) \; 10 
\end{array}
\]
\end{example}

At this point, following definitions and a few examples, most books demand from the students the task of solving more exercises. There is
no discussion as to how to master or even perform a derivation process! The very question of {\em how to prove it} is unanswered and left to the
creativeness and luck of the student. There is however a marginal, though remarkable exception in philosophy, which goes back to the
Greeks in their study of {\em analysis} and {\em synthesis}; the former depicts the backward process of working a mathematical proof, the latter concerned
with the forward derivation in a proof. More recently, a modern pioneering work in the study of {\em heuristics} is \cite{polya}, in which heuristic
strategies and guidance are given to solve mathematical problems. Some other 
proposals which provide guidance to proof construction are found in a brief discussion in \cite{hr}, and a much better one in \cite{mh}.  \\

As the above proof suggests, performing the derivation process is not easy, perhaps due to the rigidity of definition \ref{def:deriv}. Just note that more than half of the steps are instances of the $(Hyp)$ rule, which is mainly needed to be able to apply $(\to E)$.
These steps do not at all reflect a natural way of reasoning and thus make S-systems inconvenient, as opposed to Fitch-style systems. \\

The question of how to obtain such proof is quite difficult to answer directly, but a clever student can justify it by appealing to the
following mathematical reasoning. The formula we have to proof is an implication, therefore we assume the antecedent $p$ and prove the consequent $s$, which corresponds to steps 11 and 10 in the previous proof. Next we observe that $r\to s$ is part of the premises, so it suffices to prove $\G,\;p\vdash r$ (step 8). Here the application of $(\to E)$ is implicit, but the proof must be explicit and forces us to add step 9. To prove the sequent at step 8, we might try to reason as in steps 10 and 11, using the premise $q\to r$, which would lead us to seek a proof of $\G,\;p\vdash q$. But after some failed attempts we realize that $q\lor r$ is derivable from the current premises (steps 1 to 3) and that by a case analysis on this formula we can easily obtain $r$ as follows. If $q$ happens then the premise $q\to r$ yields $r$ by $(\to E)$ (steps 4 to 6), and if $r$ is the case, we are done (step 7).  Following this reasoning, the student is faced with the problem of constructing the formal proof, which requires that we start with instances of the $(Hyp)$ rule and go forward to obtain what we need. By comparison with the reasoning used to obtain the steps of the proof, this is, quite unnatural. Intuitively, all ND proofs can be constructed by such mathematical reasoning, but it is not clear in general, how to transform this reasoning into a formal derivation. An attempt, that combines both, reasoning forward from the hypotheses and backward from the conclusion, is the intercalation calculus, implemented by the {\sc AproS} system (\url{http://www.phil.cmu.edu/projects/apros/}).  Our purpose here is not to elaborate further on this, but to put forward a formal notion of backwards derivation, which we call {\sl derivation by tactics} (to be found in definition \ref{def:dertac}), one that produces proofs equivalent to the usual ones (equivalence proof will be stated as theorem \ref{thm:equiv}) and that captures the procedural reasoning {\sl ``understood''} by proof assistants.

\smallskip
Let us start by stating some heuristics. 

\subsection{Heuristics}\label{ssec:heu}

The following two strategies can be used systematically to build a proof of a given judgement:

\begin{itemize}
\item {\em Backward reasoning}: To show that $\G\vdash A$ is derivable start by analyzing the judgement and work backwards towards the axioms. To generate a proof, mantain a queue of current goals (sequents whose derivations are to be searched for), which initially includes only the original judgement. Remove a judgment from this queue, and consider a rule whose conclusion is that judgement, adding the premises of such a rule to the queue, as subgoals. This process must be repeated, with the same starting queue, for each possible rule.  The process terminates when the queue is empty, meaning all goals have been achieved.
\item {\em Forward reasoning}: To show that $\G\vdash A$ is derivable, start with axioms and work forwards towards the desired judgement. To generate a proof, extend a sequence $S$ of already derived judgements, which is initially empty, by adding to it the conclusion of any rule whose premises are in $S$.  This process generates several sequences and, assuming that all rules are considered to add new sequents, the strategy will eventually find a derivation of the original judgement. 
\end{itemize}

\noindent We are interesed in developing a goal-directed system based on backward reasoning that allows for the automation of proof construction, using a mathematical reasoning, as sketched above. To do so we restate our strategy more formally, following the proof strategies discussed in \cite{mh}.

\begin{itemize}
\item[$\bullet$]Our initial goal is the judgement we want to prove, say $\Ge\eqdef\G\vdash A$.
\item[$\bullet$]By analyzing $\G$ and $A$, we substitute the initial goal $\Ge$ by simpler subgoals, say $\Ge_1,\ldots,\Ge_k$, whose provability implies the provability of $\Ge$. After the substitution $\Ge_1$ becomes the {\sl current goal}. This process can be done using one of the following {\em heuristic rules}:
\item[$\bullet$] {\em Conclusion Analysis} (CA): The analysis of $A$ leads us to identify its 
logical form (i.e. its main operator, a connective or a quantifier) and substitute the original goal by the premises of the corresponding introduction rule. Of course, if $A$ is atomic this kind of analysis is not useful. 
\item[$\bullet$]{\em Premise Analysis} (PA): We focus on an specific formula $B$ in $\G$ which, according to our previous experience and formerly derived judgements, can help us to prove the original formula $A$. This analysis generates new subgoals either by modifying the context according to the logical form of $B$ or by modifying the conclusion. 
\item[$\bullet$]{\em Lemma Assertion} (LA): Sometimes the current goal is not directly provable by CA and PA, but follows from an intermediate judgement, like a lemma, which is derivable from the current hypotheses.   
\item[$\bullet$]A clever combination of the three previous items eventually yields a current goal which is evident (i.e. is an axiom, a hypothesis or an already proved judgement) and therefore discarded from the current sequence of subgoals.
\item[$\bullet$]The process ends succesfully when there are no more subgoals to prove. \\
\end{itemize}

A justification for the adequacy of our heuristic rules, will be given in next section.  According to the logical form of a conclusion or premise, the heuristics (CA, PA or LA) generate different subgoals. For example, if the current goal is $\G\vdash B\land C$, CA yields two subgoals, namely $\G\vdash B$ and $\G\vdash C$, being $\G\vdash B$ the new current goal. Each of these ways of generating subgoals is what we call a {\em tactic}.  \\

Even if it may seem a bit redundant, we would like to rework example \ref{ex:wex} in order to make explicit the application of our heuristic rules:
the initial goal is (the judgement in) step 11, and applying CA to this sequent yields step 10; from this PA on the premise $r\to s$ leads to step 8, thus avoiding step 9. To prove step 8, we try PA on $q\to r$, which yields $\G,p\vdash q$. This sequent is not derivable by our own means, but fortunately we realize that, according to the information on $\G$, the formula $r$ is a consequence of $q\lor r$, so we can use LA, the lemma being $q\lor r$. This is provable by means of PA on the premise $p\to q\lor r$, this avoids step 2 and yields step 1 as trivial goal, hence the lemma is proved. As we now know $q\lor r$, we can substitute the goal in step 8 by $\G,p,q\lor r\vdash r$, and then use PA on $q\lor r$, which generates two subgoals, namely steps 6 and 7, the latter is a trivial goal whereas the former is consequence of step 4 by means of PA on premise $q\to r$, and thus step 5 is avoided. At this moment, step 4 is the only remaining subgoal, which is trivial and therefore the proof is 
finished. \\

Let us present this train of reasoning in a more systematic way, one that gives guidance to the student about {\em how to prove it}. The following sequence is not a derivation in the sense of definition \ref{def:deriv}, but a sequence of backward reasonings where step $i+1$ is obtained by applying our heuristic rules CA,PA or LA to the first sequent in step $i$.
\[
\begin{array}{rll}
1 \quad & \G \vdash p\to s & \qquad \mbox{Original Goal} \\
2 \quad & \G,\;p \vdash s & \qquad  \mbox{CA.} \\
3 \quad & \G,\;p \vdash r & \qquad  \mbox{PA(}r\to s \mbox{).}
\end{array}
\]

The current goal is $\G,\;p\vdash r$. As $r$ is atomic, CA is useless; we try PA and find that $q \to r$ seems a natural candidate to continue the process, but this would lead us to a dead end when trying to prove sequent $\G,\;p \vdash q$. Instead, LA comes in action, and generates two subgoals (separated by a semicolon):
\[
\begin{array}{rll}
4\quad & \G,\;p \vdash q\lor r \;\; ; \;\; \G,\;p,\;q\lor r\;\vdash r\; & \qquad \mbox{LA(}q\lor r \mbox{).} 
\end{array} \]

\noindent The current goal is now the lemma $\G,\;p\vdash q\lor r$ and PA yields
\[ \begin{array}{rll}
5 \quad & \G,\;p \vdash p \;\;\; ;\;\;\; \G,\;p,\;q\lor r\;\vdash r\; & \qquad  \mbox{PA(} p\to q\lor r \mbox{).} 
\end{array} \]

\noindent The new current goal is $\G,\;p\vdash p$, which is trivial and can be discarded
\[ \begin{array}{rll}
6 \quad & \G,\;p,\;q\lor r\;\vdash r\;  & \qquad  \mbox{trivial}
\end{array} \]

\noindent The case analysis is now a consequence of PA($q\lor r$). The specific tactic generates two subgoals, each assuming one of $q$ and $r$. In the first case $r$ is proved by means of PA($q\to r$), and in the second the proof is trivial, for $r$ is itself a premise. This way, all subgoals are proved and the original goal succeeds.

\[ \begin{array}{rll}
7 \quad & \G,\;p,\;q\;\vdash r \;\;\; ;  \;\;\;\G,\;p,\;r\;\vdash r\; & \qquad \mbox{PA(} q\lor r\mbox{).} \\
8 \quad & \G,\;p,\;q\;\vdash q \;\;\; ;  \;\;\;\G,\;p,\;r\;\vdash r\;  & \qquad \mbox{PA(} q\to r \mbox{).} \\ 
9 \quad & \G,\;p,\;r\;\vdash r\;  & \qquad \mbox{trivial.} \\
10 \quad & & \qquad \mbox{trivial.}
\end{array}\]
The last step indicates that there are no more subgoals left and therefore the proof is finished. \\

\noindent The aim of this work is to capture the above reasoning by means of a formal system. It is clear that this kind of derivation is sequential, meaning that each step depends solely on the previous one. Therefore we can view this process as a transition system.

\subsection{A Tactical Approach to ND: Formalization}

We are now in a position to formalize the previously discussed heuristic processes by means of a transition system that manipulates sequences of goals (judgements).

\begin{definition}
  A  transition system is a triple of the form $\mathcal{T}=\langle \mathbb{S},\mathcal{F},\rhd\rangle$ where $\mathbb{S}\neq\varnothing$ is a set of states or configurations; $\mathcal{F}\subseteq\mathbb{S}$ is the set of final or terminal states; $\rhd\subseteq \mathbb{S}\times\mathbb{S}$ is a binary relation on $\mathbb{S}$, called the transition relation, such that for every $F\in\mathcal{F}$ and every $S\in\mathbb{S},\; F\not\!\rhd\; S$. That is, there are no transitions from final states. 
\end{definition}

For the sake of clarity, let us say that a goal $\Ge$ is a judgement, we denote a finite sequence of goals as $\Sc=_{def}\Ge_1;\ldots;\Ge_k$. The set of such sequences is $GSeq$, in particular we have $\square\in SeqG$, where $\square$ represents the empty goal sequence. \\

The transition system below captures the backward reasoning strategy by formalizing the process of substituting a goal with its corresponding subgoals according to the heuristic rules CA,PA or LA.

\begin{definition}\label{def:tact}
The transition system of tactics is defined as $\mathcal{T}=\langle GSeq, \{\square\},\rhd\rangle$. That is, a state is a sequence of goals; the only terminal state is the empty sequence of goals $\square$ ; and the transition relation $\rhd$ between states is given by the following rules, called {\em tactics}:
\begin{itemize}
\item Conclusion Analysis (CA):
  \begin{itemize}
		\item {\tt intro}: $\G\vdash A\to B;\Sc\;\rhd\; \G,A\vdash B;\Sc$
		\vspace{0.2cm}
		\item {\tt split}: $\G\vdash A\land B;\Sc\;\rhd\; \G\vdash A; \G\vdash B;\Sc$
		\vspace{0.2cm}
		\item {\tt left}: $\G\vdash A\lor B;\Sc\;\rhd\; \G\vdash A;\Sc$
		\vspace{0.2cm}
		\item {\tt right}: $\G\vdash A\lor B;\Sc\;\rhd\; \G\vdash B;\Sc$
		\vspace{0.2cm} 
              \item {\tt intro}: $\G\vdash \forall x A;\Sc\;\rhd\;\G\vdash A;\Sc$ where  
w.l.o.g., $x\notin FV(\G)$
	\vspace{0.2cm}
	\item {\tt exists}: $\G\vdash \exists x A;\Sc\;\rhd\;\G\vdash A[x:=t];\Sc$ 
\end{itemize}
	\vspace{0.2cm}
\item Premise Analysis (PA):
\begin{itemize}
	\item {\tt apply}: $\G,A\to B\vdash B;\Sc\;\rhd\; \G,A\to B\vdash A;\Sc$
\vspace{0.2cm}
	\item {\tt destruct}: $\G,A\land B\vdash C;\Sc\;\rhd\; \G,A,B\vdash C;\Sc$
	\vspace{0.2cm}
	\item {\tt destruct}: $\G,A\lor B\vdash C;\Sc\;\rhd\; \G,A\vdash C;\G,B\vdash C;\Sc$

	\vspace{0.2cm}
	\item {\tt destruct}: $\G,\exists x A\vdash C;\Sc\;\rhd\; \G,A\vdash C;\Sc$ 
where w.l.o.g. $x\notin FV(\G)$
\end{itemize}

\item Lemma Assertion (LA): 
\begin{itemize}
	\item {\tt assert}: $\G\vdash C;\Sc\;\rhd\; \G\vdash A;\G,A\vdash C;\Sc$
	\vspace{0.2cm}
	\item {\tt cut}: $\G\vdash C;\Sc\;\rhd\; \G\vdash A\to C;\G\vdash A;\Sc$
	\end{itemize}

\item Discarding tactics:
\begin{itemize}
	\item {\tt apply}: $\G,\forall xA\vdash A[x:=t];\Sc\;\rhd\;\Sc$
	\vspace{0.2cm}
	\item {\tt trivial}: $\G,A\vdash A;\Sc \;\rhd\; \Sc$. 
	\vspace{0.2cm}

\end{itemize}
\end{itemize}
\end{definition}

These tactics are classified according to the heuristic rules of the backward reasoning strategy: the first group is generated by each particular case of CA, according to the logical form of the conclusion; the second group comes from PA, in agreement with the logical form of the focused premise; the process of lemma assertion is given by the third group.  Observe that both LA tactics are the same from a logical point of view, but, as we are dealing with goal sequences, they are quite different from an operational point of view. Finally, discarding tactics, that is, those whose application decreases the number of subgoals, are provided by the fourth group. \\ 
We now give a brief justification of the four tactic groups: discarding tactics correspond to proving axioms, that is, sequents given by the $(Hyp)$ rule or sequents of the form  $\G,\fa x A\vdash A[x:=t]$ corresponding to universal instantiation.
LA tactics are justified by the so-called substitution property of the ND system: if $\G,A\vdash C$ and $\G\vdash A$ then $\G\vdash C$  (the formula $A$ is the lemma). LA tactics rules require the generation of a new formula ($A$), a difficult, if not an impossible task
for theorem provers to implement the transition system.
The operational mechanism of CA tactics is supported by the inversion principle\footnote{That is, the derivability of the conclusion implies 
the derivability of the premises. This property becomes a theorem when the context $\G$ is empty.} of the introduction rules. Finally, PA tactics are justified by the inversion\footnote{Again, this is a theorem for $\G=\varnothing$.} of the following admissible\footnote{That these inference rules are admissible is a direct consequence of the elimination rules (more details in the extended version of this paper).} inference rules which generate subgoals either by simplifying a premise or, in the case of implication, by modifying the conclusion: 
\[
\mspace{275mu} \frac{\G,A\to B\vdash A}{\G,A\to B\vdash B}
\]

\[
\mspace{120mu}\frac{\G,A,B\vdash C}{\G,A\land B\vdash C}\;\;\;\;\;\;\;
\frac{\G,A\vdash C\;\;\;\;\;\G,B\vdash C}{\G,A\lor B\vdash C} \;\;\;\;\;\;\;  \frac{\G,A\vdash C\;\;x\notin FV(\G,C)}{\G,\ex x A\vdash C} \;\;\;\;\;\;\;\;\;\;\;\;\;\;
\]

\smallskip
It is now clear that the backward reasoning strategy is captured by the transition system, therefore a derivation by tactics is simply a sequence of transitions ending on the final state $\square$. 

\begin{definition}\label{def:dertac}
 Let $\mathcal{J}=_{def}\G\vdash A$ be a judgement. A derivation of $\mathcal{J}$ by tactics is a sequence of states $\Sc_1,\ldots,\Sc_k$ such that $\Sc_1=\mathcal{J}$, for every $1\leq i<k,\;\Sc_i\rhd \Sc_{i+1}$ and $\Sc_k=\square$. If such a derivation exists, we write $\mathcal{J}\rhd^+\square$. 
\end{definition}

We now show a derivation by tactics of our running example.

\begin{example}
Let 
$\G=\{p \to q \lor  r ,\;q\to r,\;r\to s\}$. The following sequence of transitions
shows that $\G\vdash p\to s \rhd^+ \square$. In the right column of step $i+1$ we annotate the name of the tactic that allows for the transition from step $i$ to step $i+1$. 

\[
\begin{array}{rll}
1\quad & \G\vdash p\to s &  \\
2 \quad & \G,\;p \vdash s & \qquad \mbox{intro} \\
3 \quad & \G,\;p \vdash r & \qquad \mbox {apply} \;\; r\to s \\
4 \quad & \G,\;p\vdash q\lor r\;\;\; ; \;\;\; \G,\;p,\;q\lor r\vdash r & \qquad \mbox{assert} \;\; q \lor r \\
5 \quad & \G,\;p\vdash p\;\;\; ; \;\;\; \G,\;p,\;q\lor r\vdash r& \qquad \mbox{apply}\;\;p\to q\lor r \\
6 \quad & \G,\;p,\;q\lor r\vdash r & \qquad \mbox{trivial} \\ 
7\quad & \G,\;p,\;q \vdash r \;\;\; ; \;\;\; \G,\;p,\;r \vdash r & \qquad \mbox{destruct} \;\; q\lor r \\
8 \quad & \G,\;p,\;q \vdash q \;\;\; ; \;\;\; \G,\;p,\;r \vdash r & \qquad \mbox{apply}\;\; q\to r \\
9 \quad & \G,\;p,\;r \vdash r &\qquad \mbox{trivial} \\
10 \quad & \square & \qquad  \mbox{trivial}
\end{array}\]
\end{example}

Let us now present a more elaborated example.

\begin{example}
Let 
$\G=\{(x\lor p) \land q \to  l ,\;m\lor q\to s\land t,\;(s\land t)\land l\to x,\; p\to q\}$. The following 
is a derivation by tactics of $\G\vdash m\land p\to x$.

\[
\begin{array}{rll}
1\quad & \G\vdash m\land p\to x &  \\
2 \quad & \G,\;m\land p \vdash x & \qquad \mbox{intro} \\
3 \quad & \G,\;m\land p \vdash (s\land t)\land l & \qquad \mbox {apply} \;\; (s\land t)\land l\to x \\
4 \quad & \G,\;m, p \vdash (s\land t)\land l & \qquad \mbox {destruct} \;\; 
m\land p \\

5\quad & \G,\;m, p \vdash s\land t \;\; ; \;\; \\
&        \G,\;m, p \vdash l
 &  \qquad \mbox{split} \\ 

6\quad & \G,\;m, p \vdash m\lor q \;\; ; \;\; \\
&        \G,\;m,  p \vdash l
 &  \qquad \mbox{apply} \;\; m\lor q\to s\land t \\ 

7\quad & \G,\;m, p \vdash m \;\; ; \;\; \\
&        \G,\;m, p \vdash l
 &  \qquad \mbox{left}  \\

8\quad & \G,\;m,p \vdash l \;\; ; \;\; 
 &  \qquad \mbox{trivial} \\ 

9\quad & \G,\;m,p \vdash (x\lor p)\land q
 &  \qquad \mbox{apply}\;\;(x\lor p)\land q \to l \\ 

10\quad & \G,\;m,p \vdash x\lor p \;\; ; \;\; \\
& \G,\;m,p \vdash q  
 &  \qquad \mbox{split} \\ 

11\quad & \G,\;m,p \vdash p \;\; ; \;\; \\
& \G,\;m,p \vdash q  
 &  \qquad \mbox{right} \\ 

12\quad 
&        \G,\;m,p \vdash q
 &  \qquad \mbox{trivial} \\ 

13\quad 
&        \G,\;m,p \vdash p
 &  \qquad \mbox{apply}\;\;p\to q \\ 

14 \quad & \square & \qquad  \mbox{trivial}
\end{array}\]
\end{example}

In order to show how our approach works in first order (minimal) logic, here is an example.

\begin{example}
We give a derivation by tactics of the sequent 
\[
\vdash \forall v (Pv \rightarrow Qv) \rightarrow \forall x \big(\exists y (P y \wedge Rxy) \rightarrow \exists z (Qz \wedge Rxz)\big).
\]

 \[
 \begin{array}{rll}
 1\quad & \vdash \forall v (Pv \rightarrow Qv) \rightarrow \forall x \big(\exists y (P y \wedge Rxy) \rightarrow \exists z (Qz \wedge Rxz)\big)
 &  \\ 
 2\quad & \forall v (Pv \rightarrow Qv) \vdash \forall x \big(\exists y (P y \wedge Rxy) \rightarrow \exists z (Qz \wedge Rxz)\big)
 &  \mbox{intro} \\ 
 3\quad & \forall v (Pv \rightarrow Qv) \vdash \exists y (P y \wedge Rxy) \rightarrow \exists z (Qz \wedge Rxz)
 &  \mbox{intro} \\ 
 4\quad & \forall v (Pv \rightarrow Qv),\; \exists y (P y \wedge Rxy) \vdash  \exists z (Qz \wedge Rxz)
 &  \mbox{intro} \\ 
 5\quad & \forall v (Pv \rightarrow Qv),\; P y \wedge Rxy \vdash  \exists z (Qz \wedge Rxz)
 &  \mbox{destruct} \;\exists y (P y \wedge Rxy)\\ 
 6\quad & \forall v (Pv \rightarrow Qv),\; P y ,\; Rxy \vdash  \exists z (Qz \wedge Rxz)
 &  \mbox{destruct} \; Py \wedge Rxy\\ 
 7\quad & \forall v (Pv \rightarrow Qv),\; P y ,\; Rxy \vdash  Qy \wedge Rxy
 &  \mbox{exists} \;y \\ 
 8\quad & \forall v (Pv \rightarrow Qv),\; P y ,\; Rxy \vdash  Qy  \;\; ; \;\; \\
&          \forall v (Pv \rightarrow Qv),\; P y ,\; Rxy \vdash  Rxy 
 &  \mbox{split} \\ 
 9\quad & \forall v (Pv \rightarrow Qv),\; P y ,\; Rxy \vdash  Py \rightarrow Qy \; ;\\
&
          \forall v (Pv \rightarrow Qv),\; P y ,\; Rxy,\; Py \rightarrow Qy \vdash  Qy  \;\; ; \;\; \\
&          \forall v (Pv \rightarrow Qv),\; P y ,\; Rxy \vdash  Rxy 
 &  \mbox{assert} \;Py\rightarrow Qy \\ 
10\quad & 
          \forall v (Pv \rightarrow Qv),\; P y ,\; Rxy,\; Py \rightarrow Qy \vdash  Qy  \;\; ; \;\; \\
&          \forall v (Pv \rightarrow Qv),\; P y ,\; Rxy \vdash  Rxy 
 &  \mbox{apply} \;\forall v(Pv\rightarrow Qv) \\ 
11\quad & 
          \forall v (Pv \rightarrow Qv),\; P y ,\; Rxy,\; Py \rightarrow Qy \vdash  Py  \;\; ; \;\; \\
&          \forall v (Pv \rightarrow Qv),\; P y ,\; Rxy \vdash  Rxy 
 &  \mbox{apply} \;Py\rightarrow Qy \\ 
12\quad & 
                   \forall v (Pv \rightarrow Qv),\; P y ,\; Rxy \vdash  Rxy 
 &  \mbox{trivial} \\ 
13\quad & \square & \mbox{trivial}
 \end{array}\]
\end{example}

We end by stating the theorem that guarantees that the usual concept of derivation given in definition \ref{def:deriv}, coincides with our proposed notion of derivation by tactics given in definition \ref{def:dertac}. The proof of this theorem may be found in the extended version of this paper.

\begin{theorem}[Equivalence of $\vdash$ and $\rhd^+$]\label{thm:equiv}
  For any sequent $\J$, $\J\rhd^+\square$ if and only if $\J$ is provable.
\end{theorem}

\section{Final Remarks}
Due to lack of space we were not able to discuss {\sc Coq} properly. Nevertheless, we want to mention that its underlying first order logic mechanisms can be understood as our transition system $\mathcal{T}$. Actually, the name we give to each tactic in definition \ref{def:tact} is the name of its corresponding {\sc Coq} command (c.f. \cite{bertot}). For example, a {\sc Coq} proof of our running example is given in the following script: 

\begin{lstlisting}{h}[caption={Useless code},label=list:8-6,captionpos=t,float,abovecaptionskip=-\medskipamount]
Hypotheses (p q r s : Prop)
           (H1: p -> q \/ r) 
           (H2: q -> r)
           (H3: r -> s). 
         
Theorem Example1: p -> s.
Proof.
intro.
apply H3.
assert (q \/ r) as H4.
apply H1.
trivial.
destruct H4.
apply H2.
trivial.
trivial.
Qed.
\end{lstlisting}

The reader may realize that once the tactical approach is understood, this computer-assisted proof is quite clear. Therefore, our goal for a smooth migration from teaching to using logic has been accomplished. Indeed, our backward reasoning strategy (as discussed in \ref{ssec:heu})
has been presented to students in a computational logic course, and had
a very good reception. At the moment, we will use this material in two of
our courses (Computational Logic and Automated Reasoning), where we will test the
tactical approach as a way to introduce {\sc Coq}. The results of
this experience will be reported later. 
The main difference between our approach and that of {\sc Coq} is that in the latter hypotheses are labeled, in order to make it possible to have further reference to them. If we use these labels as variables of a $\lambda$-calculus, as is done by {\sc Coq}, we can encode the application of every inference rule with a $\lambda$-term. From this encoding, the mechanism of derivation by tactics yields a $\lambda$-term which encodes such a proof, and we can decode this proof into a usual derivation. This is the well-known Curry-Howard correspondence. Part of our future work is to explore the benefits of this powerful result for teaching logic. With respect to classical logic, we may extend our approach by allowing the use of lemmas of the form $A\lor\neg A$ (without requiring its proof), or by adding, for instance, the tactic $\G,\neg A\vdash B\; ; \; \Sc \rhd \; \G\vdash A\; ; \; \Sc$, which corresponds to proving $B$ by contradicting the hypothesis $\neg A$. However, the inversion principles that 
support our proposal are invalid in this case and therefore we cannot guarantee the feasibility of the tactical approach. A further line of research is to extend our approach to other logics. We are particularly interested in modal logic. In this respect, the second author of this paper (in \cite{l15}) has verified in {\sc Coq} the definition and properties of the ND system for modal logic developed in \cite{pd}, something which allows for a straightforward definition and implementation of our tactical approach for the case of modal logic.

\section*{Acknowledgements}

We thank the two reviewers for their comments and suggestions, which helped to improve this paper. We 
acknowledge the support of the project UNAM PAPIIT IN400514-3.


\begin{thebibliography}{}

\end{thebibliography}


\begin{thebibliography}{50}
\bibitem{bertot} Bertot Y. and Castéran P. \textsl{Interactive Theorem Proving and Program Development. Coq'Art: The Calculus of Inductive Constructions.} Texts in Theoretical Computer Science. Springer 1st ed. 2004.
\bibitem{hr} Huth M. and Ryan M. \textsl{Logic in Computer Science: Modelling and Reasoning about Systems.} Cambridge University Press. Second edition, 2004. 
\bibitem{indr} Indrzejczak. A. \textsl{Natural Deduction, Hybrid Systems and Modal Logics.} Trends in Logic. Vol. 30. Springer 2010.
\bibitem{l15} Linares-Arévalo P. S. \textsl{Deducción Natural en Lógica Modal: una implementación en Coq.} Master's thesis, Universidad Nacional Autónoma de México. 2015. 
\bibitem{mh} Manzano M. and Huertas A. \textsl {Lógica para principiantes.} Alianza Editorial, 2004.
\bibitem{pd} Pfenning F. and Davies R. \textsl{A Judgmental Reconstruction of Modal Logic.} Mathematical structures in computer science, vol. 11. 511- 540. 2001. 
\bibitem{ppl} Pierce C.B. \textsl{Lambda, the Ultimate TA: Using a Proof Assistant to Teach Programming Language Foundations.} In Proceedings  of the 14th ACM SIGPLAN International conference on Functional programming. 2009.
\bibitem{polya} Polya G. \textsl{How to solve it. A new aspect of mathematical method.} Princeton University Press. 1945. 
\bibitem{pw} Hendriks, M., Kaliszyk, C., van Raamsdonk, F., Wiedijk, F. \textsl{Teaching Logic using a state-of-the-art Proof Assistant}. 
URL: \url{http://www.cs.vu.nl/~femke/ps/formed08.pdf}.
\end{thebibliography}
\end{document}